# Anomalous diffusion in an electrolyte saturated paper matrix

Sankha Shuvra Das,[1] Sumeet Kumar,[1] Sambuddha Ghosal[2], Sandip Ghosal[3] and Suman Chakraborty[1*]

[1]*Department of Mechanical Engineering, Indian Institute of Technology Kharagpur, Kharagpur 721302, India*

[2]*Neuqua Valley High School, Naperville IL 60564, USA*

[3]*Department of Mechanical Engineering & Engineering Sciences and Applied Mathematics, Northwestern University, Evanston, IL 60208, USA*

*email: suman@mech.iitkgp.ernet.in

**ABSTRACT**

Diffusion of colored dye on water saturated paper substrates has been traditionally exploited with great skill by renowned water color artists. The same physics finds more recent practical applications in paper based diagnostic devices deploying chemicals that react with a bodily fluid yielding colorimetric signals for disease detection. During spontaneous imbibition through the tortuous pathways of a porous electrolyte saturated paper matrix, a dye molecule undergoes diffusion in a complex network of pores. The advancing front forms a strongly correlated interface that propagates diffusively but with an enhanced effective diffusivity. We measure this effective diffusivity and show that it is several orders of magnitude greater than the free solution diffusivity and has a significant dependence on the solution pH and salt concentration in the background electrolyte. We attribute this to electrically mediated interfacial interactions between the ionic species in the liquid dye and spontaneous surface charges developed at porous interfaces, and introduce a simple theory to explain this phenomenon.

**INTRODUCTION**

Flow of a liquid, often made visible by a dye, in a complex porous matrix is a very familiar phenomenon.[1-4] A classical yet elusively simple example is the diffusive spreading of a colored liquid on a piece of paper.[6] Other examples are the transfer of ink onto paper in writing,[5] the transfer of dye onto fabric in the textile industry[7] and the biochemical detection of diseases or health conditions using bodily fluids in point-of-care diagnostic procedures.[8-14]

Capillary imbibition of a liquid into a porous paper matrix is governed by the interplay of surface tension, viscous and gravity forces.[15,16] In addition, topological inhomogeneities in the porous matrix result in localized variations in the diffusive flow characteristics.[17] Nevertheless, when viewed on a macroscopic scale much larger than the scale of pore inhomogeneities, the capillary front invasion during spontaneous imbibition obeys a simple square-root dependence on time. This well-known Lucas-Washburn law[18,19] in capillaries is a consequence of the interplay of time-independence of the mean capillary pressure and a progressively increasing viscous drag in the liquid column. The same universal scaling is observed in the random porous medium of the paper matrix over length and time scales that are substantially larger than the relevant microscopic scales.[6,20,21]

Here we show that the ionic species omnipresent in colored dyes can diffuse at massively elevated speed in a paper matrix, as mediated by an intrinsic interplay of the volumetric charges in the liquid and surface charges in the paper pore via an electrokinetic route. Through experiments conducted by pre-saturating the paper pores with an electrolyte, we preclude the role of capillary pressure that is commonly attributed to possible diffusive dynamics. We find that on a scale much larger than that of the pore micro-structure, the dye spreads diffusively but with an effective diffusivity $D$ that exceeds by several orders of magnitude the free solution diffusivity $D_0$ of the dye. Our results validate reported findings on paper and fabric dyeing by exemplifying that cationic dyes are known to have improved spreading and adsorption characteristics.[7] We further offer a simple explanation that substantiates our proposition on the underlying electrokinetic route by establishing the sensitivity of the spreading behavior on the salt concentration and pH of the solution.

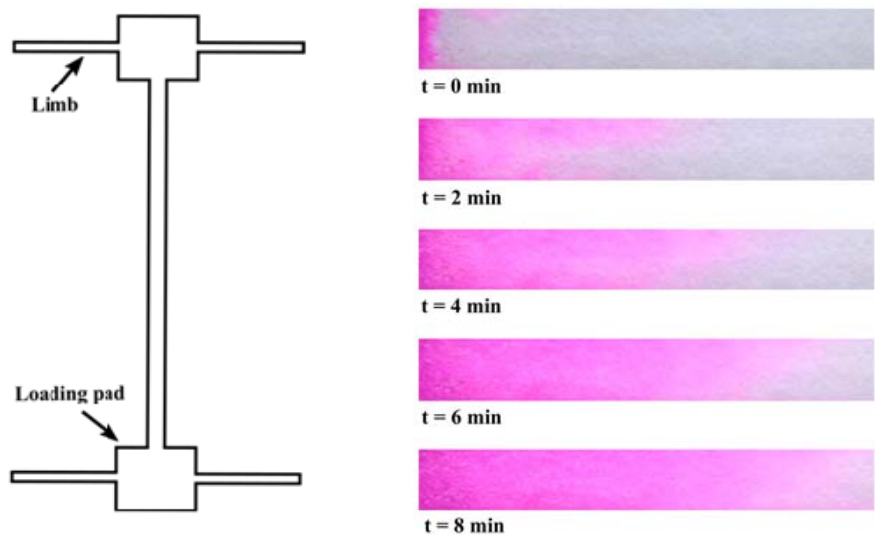

**Figure 1.** Schematic diagram showing the experimental set up (left) and frames from the movie showing discussion of dye along the paper strip (right).

**EXPERIMENTAL SECTION**

**Device fabrication and experimentation.** In order to study the diffusion of dye on a paper matrix, we cut paper strips (5 mm width and 5 cm long) from cellulose filter paper (Whatman, Grade 4, mean pore diameter ~ 20 − 25 μm, paper thickness ~ 100 μm). At either end of the paper strip there are 6×8 mm² rectangular loading pads. These are further connected to two thin segments or 'limbs' 2 cm in length and 2 mm width at either end of the loading pads (see Figure 1). To prime the strip, the limbs are immersed in external reservoirs filled with stock electrolyte solution (back ground electrolyte, BGE) for one hour. Furthermore, immediately before the start of the experiment, the paper strip is completely saturated with the BGE to ensure that any capillarity driven fluid flow is eliminated. The BGE is prepared by dissolving crystalline KCl (Merck Life Science) solid in 18.2 MΩ-cm DI water (Millipore India) at concentrations in the range 0.01 mM to 1.0 M. The test solution is prepared by mixing Rhodamine B dye (Sigma-Aldrich) in the BGE at a concentration of ~1 gm/L. In the control experiment, we use only DI water. In the investigation of the effect of pH, we added standard solutions of either $H_2SO_4$ or NaOH to DI water to successively vary the pH in the range 2.1 to 11.3. The pH was measured using a pH-meter (Oakton, Multi parameter PCSTestrTM 35). In each experiment, a constant volume of 80 μL of the test solution is transferred by micropipette to the inlet loading pad of the wetted paper strip. In order to prevent overflow at the loading pads, the solution is dispensed in small increments, approximately once every two minutes, over the duration of the experiment. The test solution diffuses along the length of the paper while the concentration at the loading pad remains approximately constant. The progression of the dye is recorded by a digital camera (Nikon D5200) in video mode at 50 fps frame rate and 1920×1080 pixel resolution (an example of such a movie is included in the accompanying Supplementary Information). Still frames at approximately 1 minute intervals are extracted from the movie footage for analysis (see Figure 1).

The images are subsequently processed using an in-house Matlab code. In order to minimize edge effects a rectangular strip 890 × 121 pixels along the centerline of each frame is selected and the grayscale intensity is averaged over the width of the strip yielding the mean intensity profile at each time. For the purpose of our analysis, we assume that this intensity is proportional to the dye concentration. In order to verify this, we conducted a separate calibration experiment where known concentrations of dye were placed on the paper and the corresponding intensity recorded. It was observed that (see Figure S1, in Supplementary Information file) the intensity was proportional to the concentration up to a limiting concentration of about 1 gm/L. The concentration in the experiments was always maintained at or below this limiting concentration. A second source of error in this experiment is the irreversible adsorption of the dye onto the paper fibers, particularly since the dye and fibers have opposite charge. A mathematical treatment of this would require us to partition the dye concentration into dissolved and adsorbed phases thereby introducing additional parameters that would preclude the determination of $D$ by the simple method presented here. We avoid this by taking a relatively high concentration of the dye (but not exceeding the limiting concentration mentioned above) so that the adsorbed concentration is small compared to the dissolved concentration. Furthermore, since irreversible adsorption is a monotonically increasing function of time, we keep the total duration of the experimental runs below 20 minutes. Irreversible adsorption is signalled by the appearance of a time-independent plateau region in the concentration profile and care must be taken to keep the concentration at the loading pad sufficiently high and the total run time sufficiently short to avoid this regime.

**RESULTS AND DISCUSSION**

We assume that the dye concentration $c(x,t)$ along the paper obeys the diffusion equation with effective diffusivity $D$. The loading pad may be considered a reservoir holding a fixed concentration of dye $c_m$. Thus, $c(0,t) = c_m$ where $x = 0$ marks the boundary between the paper strip and the inlet reservoir pad. The time dependent concentration profile is then given by,

$$c(x, t) = c_m \left[ 1 - erf\left( \frac{x}{2\sqrt{Dt}} \right) \right], \qquad (1)$$

where $erf(z) = \frac{2}{\sqrt{\pi}} \int_0^z \exp(-\xi^2) d\xi$ is the error function. From each of the frames, the location $x = x_*$ is extracted where $c(x,t)$ is exactly $c_m/2$. In view of the proportionality between the concentration and intensity, this corresponds to the location on the paper strip where the intensity is half of the maximum intensity. From Eq. (1), $x_* = 2z_*\sqrt{Dt}$ where $z_*$ is defined by $erf(z_*) = \frac{1}{2}$. We verify this relationship by plotting $x_*^2$ against time t which gives the expected proportional relationship (Figure 2). The slope of the straight line is $4z_*^2 D$ from which $D$ is determined. To further verify the diffusive nature of the transport, we rewrite eq. (1) in dimensionless form as:

$$c/c_m = 1 - erf(z_* x / x_*), \qquad (2)$$

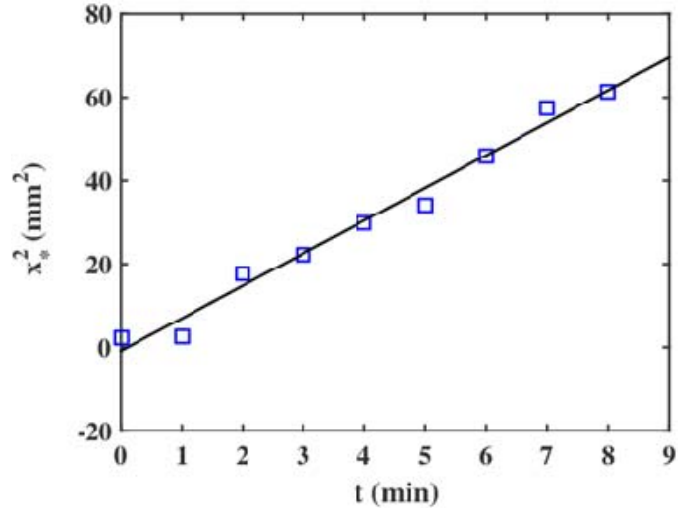

**Figure 2.** Plot of $x_*^2$, where $x_*$ is the location where the concentration is half the reservoir concentration, against elapsed time t showing the proportionality expected for diffusive spreading (BGE 1 mM KCl pH = 6.75).

Figure 3 shows the time progression of the measured concentration profiles and the inset shows the same data plotted in dimensionless form: $c/c_m$ as a function of $x/x_*$. The solid line is the theoretical curve Eq. (2). The concentration profiles at different times are seen to exhibit self-similar collapse onto the theoretical profile Eq. (2) to within expected experimental errors. The main source of error is the inherent noise in the data due to the heterogeneous nature of the paper matrix. Two of the profiles show a large deviation from the theoretical curve for large values of $x/x_*$. These profiles correspond to short times, where due to the sharp step-like variation of the concentration profile measurement is not accurate.

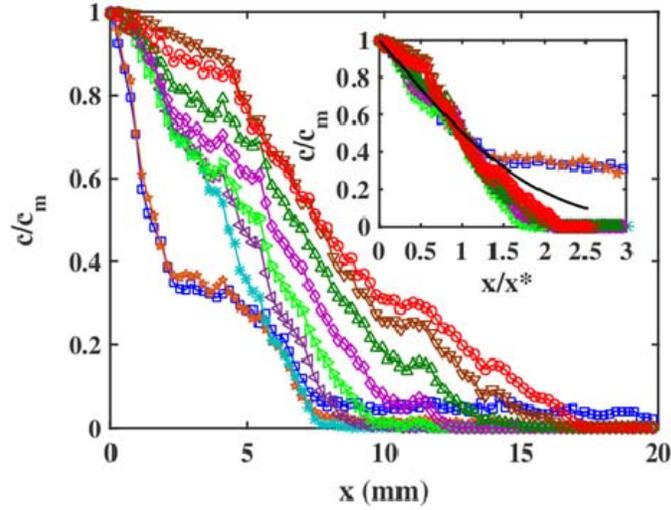

**Figure 3.** Relative concentration distribution of dye at 1 minute intervals (BGE 1 mM KCl pH 6.75). Inset shows the same data plotted against $x/x_*$. There is self-similar collapse onto the theoretical profile (solid black line) given by eq. (2).

The experimental protocol described above is repeated with the same dye solution but different salt concentrations. For each BGE the Debye length, $\lambda$ was calculated using the formula $\lambda^{-2} = \sum z^2 e^2 n / \varepsilon kT$ where $e$ is the electronic charge, $z$ is the ion valence, $n$ is the ion concentration, $\varepsilon$ is the permittivity of water, $kT$ is the Boltzmann temperature and the summation is over all ion species, which in this case consists of $K^+$ and $Cl^-$ together with the hydrogen and hydroxyl ions from the auto-dissociation of water. Figure 4 shows how the measured diffusivity varies with the Debye length. Each data point is an average over $N = 10$ separate determinations of the diffusivity and the error bars show the standard error: $\sigma / \sqrt{N}$, where $\sigma$ is the standard deviation, of the measurements. The measured pH for each BGE is also shown. The effective diffusivity is found to be about two orders of magnitude higher than the free solution diffusivity[22] of Rhodamine B: $D_0 = 0.44 \times 10^{-9}$ m²/s. The diffusivity is seen to increase with the Debye length ($\lambda$) to $\lambda \sim 10$ nm but then saturates and becomes essentially independent of the Debye length. Figure 5 shows how the measured diffusivity varies with the pH of the BGE in the absence of any added KCl. The diffusivity varies by about a factor of two and appears to show a peak at about pH 9.

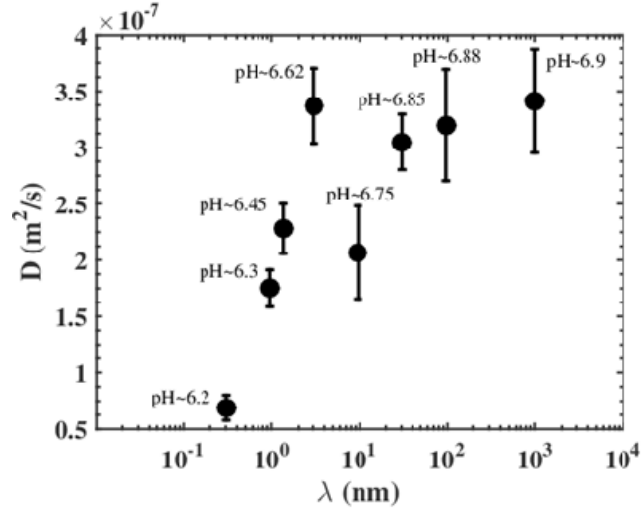

**Figure 4.** Measured effective diffusivity as a function of the Debye length. Symbols indicate an average of 10 separate diffusivity measurements and the error bars indicate the corresponding standard error. The rightmost data point corresponds to DI water and the remaining data points were obtained by progressively increasing the KCl concentration up to a maximum concentration of 1 M. The pH changes slightly on changing the KCl concentration and the measured pH for each BGE is noted next to the data point.

One may expect the diffusivity of a dye in a porous net-work to be smaller than in free solution due to decreased mobility of the solute molecules.[23] The fact that it is two orders of magnitude larger in the paper matrix is quite surprising and requires an explanation. The dependence of the diffusivity on the Debye length and solution pH (which determines the surface charge of dye and paper fibers) suggests that the underlying mechanism may be electrokinetic in nature. Figure 6 shows SEM images of the filter paper at a low resolution of ×300 as well as at a much higher resolution ×30, 000. The paper matrix is seen to have pores on a large range of scales from tens of microns to tens of nanometers. At the lower end of this continuum of length scales, screened Coulomb interactions between the cationic dye and the anionic pores could significantly speed up diffusion. Indeed, under these circumstances, the Gibbs-Donnan effect[24] would create an electric field directed into the pores that would electrophoretically drive the charged dye into these gaps. In general, the diffusivity $D \sim (\Delta u)l$, where $\Delta u$ is a characteristic velocity fluctuation scale and $l$ is a length scale. When the dye diffuses in free solution $\Delta u \sim \sqrt{kT/m}$, where $m$ is the molecular mass of the dye and $l$ is the mean free path. When the molecule is electrophoretically driven into narrow gaps there is an additional velocity scale $u_{ep}$, the characteristic electrophoretic velocity of the dye molecule and an additional length scale $\lambda$, the Debye shielding length, which is much larger than molecular mean free paths. Another possible electrokinetic mechanism is diffusiophoresis.[25,26] The presence of the dye creates a concentration gradient along the paper strip. Such a gradient, in the presence of a charged substrate (the paper fibers) could result in a diffusiophoretic flow which may appear as an anomalously fast diffusion.

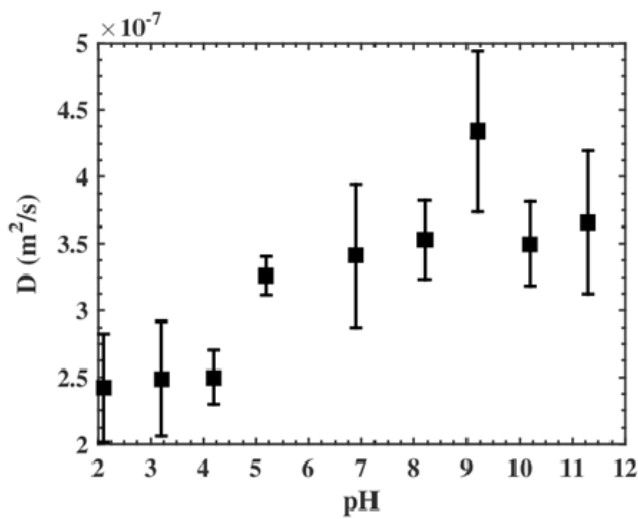

**Figure 5.** Measured effective diffusivity as a function of pH at zero KCl concentration. Symbols indicate an average of 10 separate diffusivity measurements and the error bars indicate the corresponding standard error.

Paper-based diagnostic devices that rely on the movement of reagents and bodily fluids through a fibrous network and often use color changes to signal antigen- antibody reactions in functionalized substrates are of great interest in the purview of affordable healthcare. An understanding of the basic physics of molecular transport in paper media is therefore of utilitarian implication. Research on this topic to date has largely been focused on the capillarity based movement of fluid described by the Lucas-Washburn law. However, advection of a species by capillary driven flow is also accompanied by diffusive spreading. The relative importance of the two is governed by the P´eclet number $Pe = uw/D$, where $u$ is a characteristic flow speed, $w$ is the paper width and $D$ is the effective diffusivity. Diffusive processes are important if $Pe \sim 1$ or $u \sim D/w$. If we use our measured values of diffusivity and assume that the paper width is of the order of millimeters, we obtain $u \sim$ mm/s which is fairly typical of the imbibition of capillarity fronts through paper strips. Thus, in studying transport through paper, an understanding of the diffusive processes is just as important as understanding capillarity driven transport.

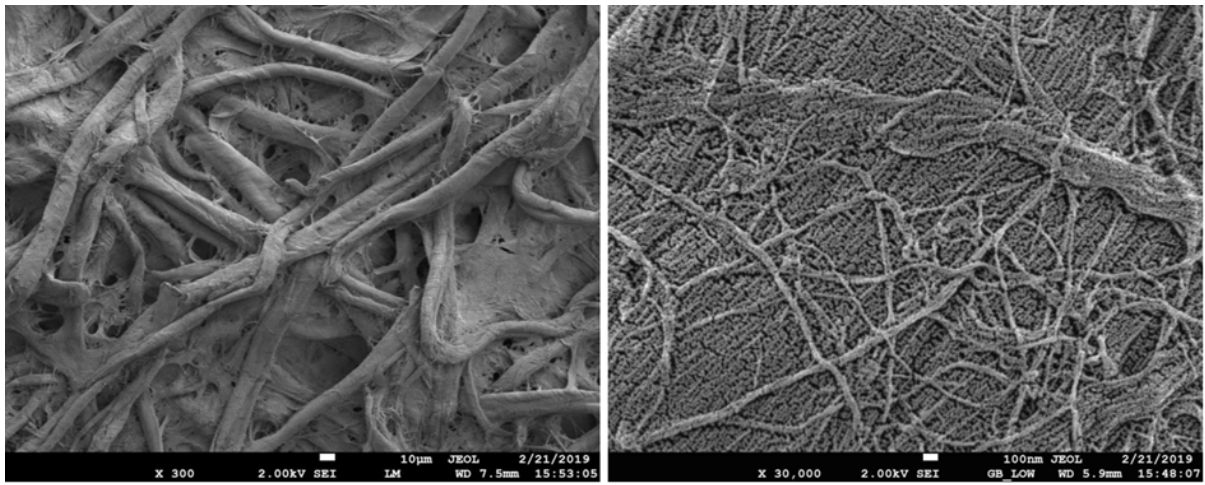

**Figure 6.** SEM images of the surface of Whatman Grade 4 filter paper at × 300 magnification (left) and × 30, 000 magnification (right). The scale bars represent 10 μm and 100 nm on the low and high resolution images respectively. The presence of a 'fractal-like' structure is seen with pores down to the 1 − 10 nm range.

**CONCLUSIONS**

In this study, the paper was pre-saturated with fluid so that capillarity was unimportant. We found that though the transport on scales much larger than the pore dimensions obeys diffusive scaling, the effective diffusivity is about two orders of magnitude higher than the free solution diffusivity of Rhodamine B. We hypothesized that this enhancement of the diffusivity is due to electrokinetic effects arising from the chemical charge of the cationic dye and paper fibers. This is supported by the observed dependence of the measured diffusivity on the salt concentration of the background electrolyte and pH of the solution. The former controls the Debye shielding length and the latter the charge state of the dye and paper fibers. In reality, however, additional complications may arise due to a continuous swelling of the paper medium along with a progressive deposition of particulate matters from the body fluid or the colored dye. Understanding the complications of combining the dynamics of swelling and electrokinetically driven imbibition in a topographically complicated porous matrix with the rate of deposition at a location for on-spot diagnostics may create new vistas for developing functionalized paper substrates for more accurate quantification of diseased conditions using body fluids in resource-challenged environments.

## ACKNOWLEDGEMENTS

SG2 (author 4) & SC acknowledge Institute Challenge Grant (SGCIR) from IIT Kharagpur & support from Indo-US joint center on "Nanoscale Transport & Biological Interfaces" from IUSSTF. SC acknowledges DST, Govt. of India, for J. C. Bose National Fellowship. SG2 acknowledges support from USIEF (Award no. 7428). SG1 (author 3) and SG2 thank IIT Kharagpur for hospitality.

# Electronic Supplementary Information

# Anomalous diffusion in an electrolyte saturated paper matrix

Sankha Shuvra Das,[1] Sumeet Kumar,[1] Sambuddha Ghosal[2], Sandip Ghosal[3] and Suman Chakraborty[1*]

[1]*Department of Mechanical Engineering, Indian Institute of Technology Kharagpur, Kharagpur 721302, India*

[2]*Neuqua Valley High School, Naperville IL 60564, USA*

[3]*Department of Mechanical Engineering & Engineering Sciences and Applied Mathematics, Northwestern University, Evanston, IL 60208, USA*

*email: suman@mech.iitkgp.ernet.in

**1. Calibration Experiment:**

In order to obtain an optimal dye concentration and to avoid saturation of dye intensity, we performed a calibration experiment. The sample solutions were prepared by mixing Rhodamine B in 1 mM KCl solution in an appropriate dye concentration ratio. The dye concentration was varied from 0.25 gm/L to 3 gm/L and the corresponding grey scale intensity was recorded. Fig. 1 shows the recorded intensity as a function of the known dye concentration. It is observed that a proportional relationship between these two quantities exists up to a dye concentration of about 1 gm/L, beyond which the intensity saturates and becomes insensitive to any further increase of dye concentration. Since in this experiment, the gray scale intensity of the video image is used as a proxy for the actual dye concentration, we take care to use as high a dye concentration as possible but not exceeding this threshold for saturation.

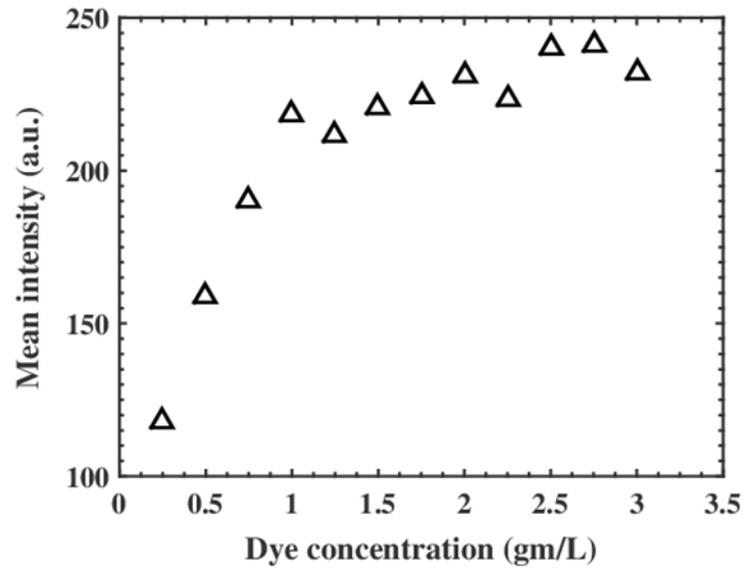

**Figure S1.** Calibration data showing gray-scale mean intensity of digital image (arbitrary units) vs. Rhodamine B dye concentration (gm/L).